\documentclass[prb,showpacs,preprint,floatfix]{revtex4}
\usepackage{graphicx}
\usepackage{bm}
\usepackage{epsf}

\begin{document}

\title{Quantum hydrogen vibrational dynamics in LiH: new neutron measurements and 
variational Monte Carlo simulations}
\author{J. Boronat$^{1}$, C. Cazorla$^{1}$, D. Colognesi$^{2}$, and M. Zoppi$^{2}$}
\affiliation{
$^{1}$ Departament de F\'{\i}sica i Enginyeria Nuclear, 
Campus Nord B4-B5, Universitat Polit\`ecnica de Catalunya, 
08034 Barcelona, Spain \\}
\affiliation{
$^{2}$ Consiglio Nazionale delle Ricerche, Istituto di Fisica Applicata
'Nello Carrara', via Madonna del Piano, 50019 Sesto F.no (FI), Italy}

\date{\today}


\begin{abstract}
Hydrogen single-particle dynamics in solid LiH at $T=20$ K has been
studied through the incoherent inelastic neutron scattering technique. A
careful analysis  of the scattering data has allowed for the determination
of a reliable hydrogen-projected  density of phonon states and, from this,
of three relevant physical quantities: mean squared displacement, mean
kinetic energy, and Einstein frequency. In order to interpret  these
experimental findings, a fully-quantum microscopic calculation has been
carried out  using the variational Monte Carlo method. The agreement
achieved between neutron scattering data and Monte Carlo estimates is
good.  In addition, a purely harmonic calculation has been also performed
via the same Monte Carlo code, but anharmonic effects in H dynamics were not found
relevant. The possible limitations of the present semi-empirical
potentials are finally discussed.
\end{abstract}


\maketitle

\section{Introduction}
Hydrogen forms stable stoichiometric hydrides by reaction with all of the
alkali metals: Li, Na, K, Rb, and Cs.\cite{Mueller} Since the first x-ray
study \cite{Zintl} in 1931, it has been shown that LiH, NaH, KH, RbH and
CsH (AlkH in short) crystallize with the rock-salt structure 
(i.e. a fcc with a two-atom basis: Li in $(0,0,0)$ and H shifted by  
$(a/2) ({\bf i} + {\bf j} + {\bf k})$) at room temperature. 
In these materials, experimental evidence seems  consistent with hydrogen
being present in the form of anions or modified anions: electron
distribution investigations  \cite{Calder} estimated the ionic charge in
LiH to fall in range from 0.4 to 1.0 electron-charges, indicating that
the alkali hydrides are probably very similar to the alkali halides, with
respect to the electronic structure, and that  AlkH might even be regarded
as the lightest alkali halides, not far from alkaline fluoride compounds:
AlkF. In addition, calculations of the electronic charge density in LiH, 
based on the local density approximation, suggest that most of the electronic 
charge is transferred from Li to H.\cite{Rodriguez} Because of this fact, 
which gives rise to long-range interactions
between hydrogen atoms, the H dynamics in these  compounds is expected to
be very different from the one in group III-VIII metal hydrides. However,
apparently only LiH has been studied in some detail,
\cite{Islam} both theoretically and experimentally. 
\par
The choice of lithium hydride (and deuteride) has not been casual at all:
they are rock-salt crystals having only  four electrons per unit cell,
which makes them the simplest ionic crystals in terms of electronic
structure. Moreover, there is a large isotopic effect provided by the
substitution of the proton with the deuteron.
Last but not least, owing to the low  mass of their
constituent atoms, these two compounds represent a good case for the
calculations of the zero-point motion  contribution to the lattice
energy.  In this respect, they are expected to behave as
{\it quantum crystals}, somehow
related to the hydrogen (H$_2$ and D$_2$) \cite{VanK} or the helium ($^3$He
and $^4$He) \cite{Glyde} families. By comparing an important anharmonicity
parameter, namely the zero-temperature Lindemann ratio \cite{Lin}
to the ones of other quantum crystals, one
finds that H in LiH ($\gamma_{H}=0.12$)\cite{Vid} lies in between solid 
H$_2$ ($\gamma=0.18$) and solid
Ne ($\gamma=0.09$), placed  very close to solid D$_2$ ($\gamma=0.14$).  
Because of these unique physical properties (and incidentally because of 
their use in thermonuclear weapons) LiH and LiD are in general fairly well 
described, even in condition of very high pressure  ($p>10$ GPa): both 
structurally (neutron diffraction),\cite{Besson} and dynamically 
(second-order Raman spectroscopy),\cite{Ho} and through ab-initio 
electronic structure simulations.\cite{Nova}  
\par
However, as far as complete (i.e. including full
phonon dispersion curves) lattice dynamics works are concerned, the
situation looks much less exhaustive: LiD dispersion curves have been
measured by coherent inelastic  neutron scattering in 1968,\cite{Verble}
even though the longitudinal optical branch  was almost unobserved. These
experimental data were fitted through a 7-parameter shell model (SM7) and
converted into equivalent LiH data. Later on,\cite{Dyck} the same data
have been fitted again through a more advanced method:  a deformation
dipole model with 13 adjustable parameters (DDM13), which also provided
elastic and dielectric  constants, effective charges, second-order Raman
spectra, etc., all in good (or fair) agreement with the known  experimental
values. The only noticeable exception was represented by the H-projected
density of phonon states (H-DoPS) for LiH:\cite{Izyumov}
\begin{equation}
\label{defproj}
Z_{\rm H}(\omega)=\frac{1}{3N} \sum_{{\bf q}} \sum_{j=1}^6
\left| {\bf e}_{\rm H} ({\bf q},j)\right|^2
\delta(\omega-\omega({\bf q},j))~,
\end{equation}
where ${\bf q}$ is a phonon wave-vector contained in the first Brillouin zone, 
$N$ is the number of these wave-vectors, 
$j$ is labeling the six phonon branches, 
${\bf e}_{\rm H} ({\bf q},j)$ is the polarization vector for H, 
and $\omega({\bf q},j)$ is the phonon frequency.
LiH H-DoPS turned out both from SM7 and DDM13 to be rather different from the old incoherent inelastic
neutron scattering (IINS) measurement.\cite{Zemlianov} In addition, both SM7
and DDM13 are force constant models and do not provide  any suitable
inter-ionic potential scheme for LiH. The potential approach was first
attempted by Hussain and Sangster,\cite{Huss} who tried to include
alkali hydrides in a larger potential scheme derived for all the
alkali halides making use of the Born-Mayer functional form: considering
the little number (i.e. 3) of adjustable parameters, the LiD dispersion
curves obtained from this potential are quite in good agreement with the coherent
neutron scattering data.\cite{Verble} However, the existence of some
problems in this scheme for LiH (e.g. the H-H Pauling parameter turned
out to be unphysical, the elastic constants were too large, etc.),
prompted Haque and Islam \cite{Haque} to devise a new set of potentials
(HI) for both LiH and NaH. But, despite their better description of some of
the macroscopic properties of LiH and their superior physical soundness, the
HI potential does not provide any advantage over the older one as far as the 
dispersion curves are concerned. Finally, it is worth mentioning an important
ab-initio calculation of the LiD dispersion curves,\cite{Bert} which
correctly included the zero-point effects in the lattice energy
minimization procedure.
\par
As for thermodynamics, accurate constant-pressure specific heat
measurements on LiH and LiD  in the temperature range $T=(4-300)$ K are
reported by Yates and co-workers.\cite{Yates}  These experimental
results, once corrected for thermal expansion with the help of temperature
dependent lattice  parameters,\cite{Anderson} were used to extract the
effective Debye temperature, $\Theta_D$, as a function of  $T$ in the
aforementioned temperature range.  The relative thermal variation of
$\Theta_D$ in LiH was found to be very large: 20\% between $4$ K and
$80$ K;  more than 30\% between $80$ K and room temperature, revealing a
strong anharmonicity in the LiH lattice dynamics.  Later, a comparison was
made between experimental $\Theta_D(T)$ data and the DDM13 results of Dyck
and Jex.\cite{Dyck} The agreement was found quite good between $30$ K and
$300$ K, while for temperature values below  $20$ K some anomalies in the
behavior of the experimental specific heat were found, i.e., the measured 
heat capacity displays a peak at $T_c=11.1$ K (for LiH), the origin of
which has not yet been discovered.  The possibility of a phase transition
to a CsCl-type structure has been proposed,\cite{Yates} but it seems
rather  unlikely. The specific heat measurements in the reported
temperature range are mainly sensitive to the details  of the phonon
density of states in its low-energy region (say $0-30$ meV),
\cite{Wallace} which lies for LiH in  the acoustic band, spanning the
phonon energies in the $(0-70)$ meV range. Thus, unfortunately,
thermodynamics seems  rather unable to probe the real H dynamics in LiH. 
\par
Given the aforementioned scenario, a study on the hydrogen vibrational
dynamics in LiH has to answer a number of questions that we have tried to
synthesize in four main points: 1) How is the H vibrational dynamics in LiH
described  by the proposed force-constant and potential schemes? 2) How do
they compare with new and reliable incoherent inelastic neutron data? 
3) Are low-temperature anharmonic effects (i.e., purely quantum-crystal effects)
detectable in the H  dynamics in LiH? 4) In case, how is it possible to
calculate them using microscopic theory?  
We believe that a combined use of incoherent inelastic neutron
scattering from which the H-DoPS can be worked out, and of fully quantum
simulations, from which important equilibrium physical quantities can be
derived (mean squared displacement, mean kinetic energy and Einstein
frequency) can provide a new and deep insight into the problem of quantum
hydrogen dynamics in condensed matter. The rest of the present work will
be developed according to the following scheme: the neutron measurements
will be described in Sect. II, and it will be also shown how to extract a
reliable H-DoPS from the experimental spectra. Sect. III  will be
devoted to describe the simulation technique, namely the variational Monte Carlo
method. Then, in Sect. IV the
results on the H dynamics in LiH will be discussed. Here a comparison between
the quantities derived from the experimental spectra and their estimates
obtained through the variational Monte Carlo simulations, and  other less
advanced techniques, will be established. Finally, in Sect. V  we will draw 
some conclusions.
%
\section{Experimental and data analysis}
The present neutron scattering experiment was carried out on TOSCA-II, a
crystal-analyzer inverse-geometry spectrometer \cite{Tosca} operating at
the ISIS pulsed neutron source (Rutherford Appleton Laboratory, Chilton,
Didcot, UK).  The incident neutron beam spanned a broad energy ($E$) range
and the energy selection was carried out on the secondary  neutron
flight-path using the (002) Bragg reflection of 7 graphite single-crystals
placed in back-scattering  around $137.7^{\rm o}$.  This arrangement fixed
the nominal scattered neutron energy to $E^\prime=3.32$ meV.  Higher-order
Bragg reflections were filtered out by 120 mm-thick beryllium blocks
cooled down to a temperature  lower than 30 K. This geometry allows to
cover an extended energy transfer ($3$ meV$<\hbar \omega<500$ meV) range, 
even though the fixed position of the crystal analyzers and the small
value of the final neutron energy imply a  simultaneous variation in the
wave-vector transfer, $Q$.  In other words, on TOSCA-II $Q$ is a monotonic
function of $\omega$: 2.8 \AA$^{-1}$  $<Q(\omega) \sim \omega^{1/2}<16.5$
\AA$^{-1}$.  The resolving power of TOSCA-II is rather good in the
accessible energy transfer range: $\Delta \hbar \omega/E \simeq 1.3-2.3$\%.
\par
The sample cell was made of aluminum (0.8 mm-thick walls) with a
squared-slab geometry, exhibiting an internal gap  of $1.2$ mm. The cell
area (85.0 mm $\times$ 85.0 mm) was rather larger than the neutron beam
cross-section  (squared, roughly 40 mm $\times$ 40 mm). The scattering sample
was made of 5.733 g of polycrystalline LiH (powder from Sigma-Aldrich,
97\% assay) and  contained purely natural lithium (92.5\% $^7$Li and
7.5\% $^6$Li). After collecting background data on the empty cryostat,
we cooled down the empty sample cell to the experimental temperature
($T$=20 K), starting to record the time-of-flight (TOF) neutron spectrum
as $T<30$ K, up to an integrated proton current of 344.1 $\mu$A h.  Then
the system (LiH sample + can) was loaded into the cryostat, cooled down to 20.0
K and measured up to an integrated proton current of 2101.0 $\mu$A h. The
thermal stability of this measurement was quite satisfactory, since the
temperature fluctuations never exceeded 0.3 K, and the temperature gradient
between top and bottom of the cell was lower than 0.1 K. The mean sample
temperature was estimated to be ($20.1 \pm 0.1$) K. Special care was
devoted to prevent possible lithium hydroxide formation during the sample
loading procedure. In addition, the subsequent inspection of the raw
spectroscopic data exhibited no sharp features in the (75-78) meV range,
which are a clear signature of lithium hydroxide contamination.\cite{Park} 
\par
The raw neutron
spectrum exhibited a series of strong features (in the $73-140$ meV range)
related to the density of optical-phonon states and essentially due to the
H and Li anti-phase motion in the lattice unit cell. The first overtones
of  these optical bands are also well visible in the recorded spectrum at
about twice the aforementioned energy transfer intervals. On the
contrary, at low energy transfer $\hbar \omega$, the acoustic band appears
rather weaker than the optical one, since here lithium and hydrogen ions
move basically in phase, and the H mean square displacement is much
smaller, more than 20 times according to lattice dynamics simulations
\cite{Dyck}.
\par
\begin{figure}
\begin{center}
\epsfxsize=30pc
\epsfbox{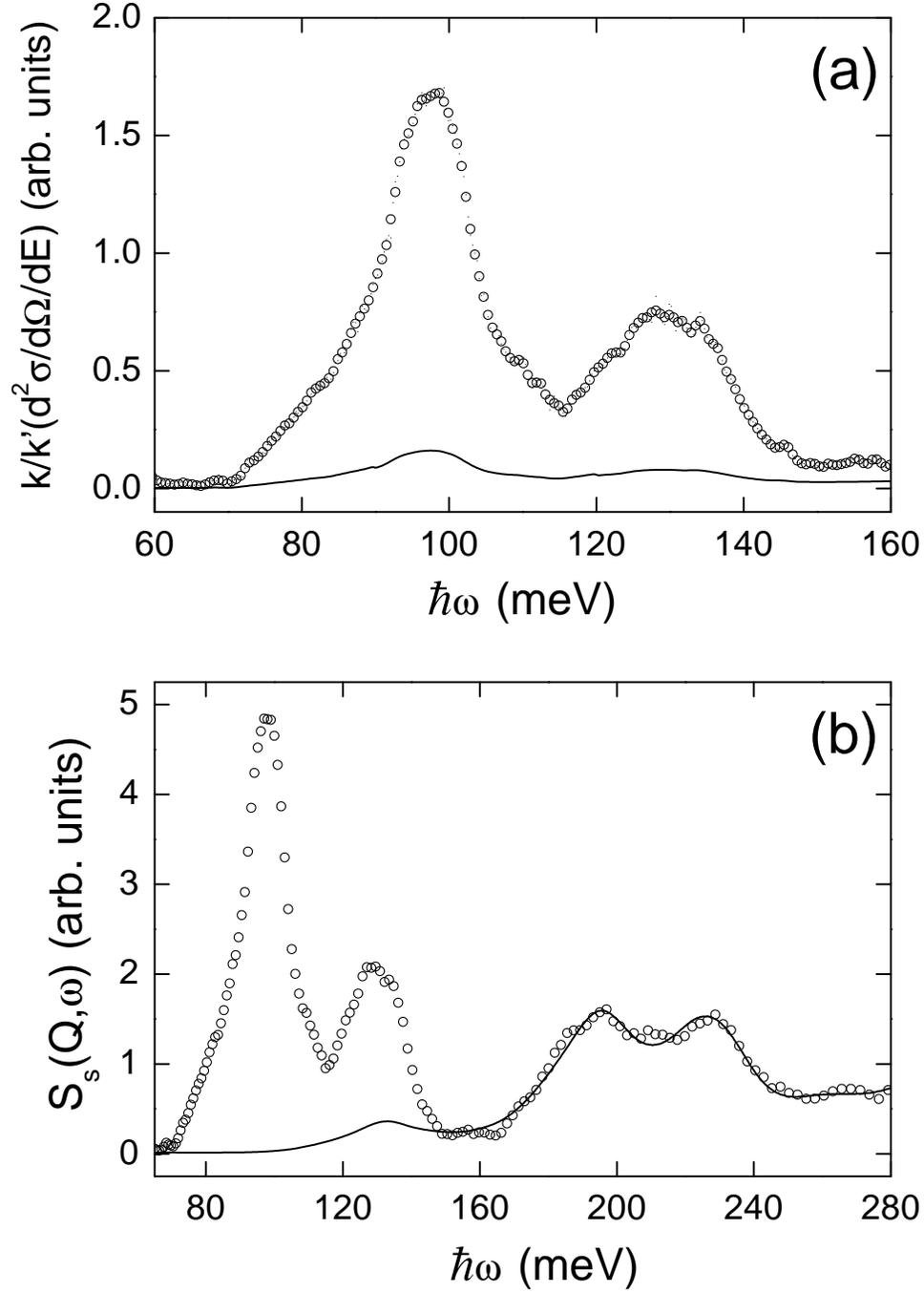}
\end{center}
\caption{
Neutron scattering spectra from LiH at $T=20.1(1)$ K:
(a) TOSCA-II experimental data (circles) together with the
estimate of the multiple scattering contribution (line);
(b) experimental scattering law (circles) and its multiphonon 
component (line). 
}  
\label{MULT}
\end{figure}

The experimental TOF spectra were transformed into energy transfer data,
detector by detector, making use  of the standard TOSCA-II routines
available on the spectrometer, and then added together in a single block.
This procedure  was justified by the narrow angular range spanned by the
detectors, since the corresponding full-width-at-half-maximum,  $\Delta
\theta$, was estimated to be only $8.32^\circ$ (see Ref. [\onlinecite{Tosca}]). 
In this way,
we produced a double-differential cross-section measurement along the
TOSCA-II kinematic path  $(Q(\omega),\omega)$ of the  (LiH+ can)
system, plus, of course, background and empty can.  Then, data were
corrected for the $k^\prime/k$ factor, and background and empty-can
contributions  were properly subtracted. At this stage the important
corrections for the {\it self-absorption} attenuation and {\it multiple
scattering} contamination were performed, the former being particularly 
important due to the large natural Li absorption cross-section
($\sigma_{ab}({\rm Li})= 70.5$ barn).\cite{Love} 
These two corrections were applied
to the experimental data through the analytical approach suggested by
Agrawal in  the case of a flat slab-like sample.\cite{Agra} We
made use of the simulated H- and Li-projected densities of phonon states
derived by Dyck and Jex \cite{Dyck} in order to evaluate: 1) the H total
scattering  cross-section, known to be largely dependent on $\omega$; 2)
the LiH scattering law to be folded on itself in order to generate the
multiple scattering contributions. Both procedures were accomplished in
the framework of the incoherent approximation,\cite{Love}  totally
justified by the preponderance of scattering from H ions, and by the
polycrystalline nature  of the sample. Multiple scattering was found to be
around 6.7\% of the total scattering in the energy transfer range of
main interest (i.e. optical phonon region):  $50$ meV$<\hbar \omega<150$ meV,
and then subtracted as shown in Fig. \ref{MULT} (a). After performing the two
aforementioned corrections, our neutron spectrum still contained some
scattering  from the Li ions, which was however much lower than the one
from the H ions, and, moreover, mainly  localized in the acoustic phonon
region,\cite{Dyck,Zemlianov} i.e. for  $\hbar \omega<50$ meV. So the Li
contribution was simulated (always in the framework of the incoherent
approximation) through the Dyck and Jex calculations of the Li-projected
densities of phonon states \cite{Dyck} and then removed. All the practical
details of this procedure can be found in Ref. [\onlinecite{Colo}], where
this is applied to various binary solid systems, such as H$_2$S, D$_2$S,
and HCl, measured on TOSCA-I.  
\par
\begin{figure}
\begin{center}
\epsfxsize=28pc
\epsfbox{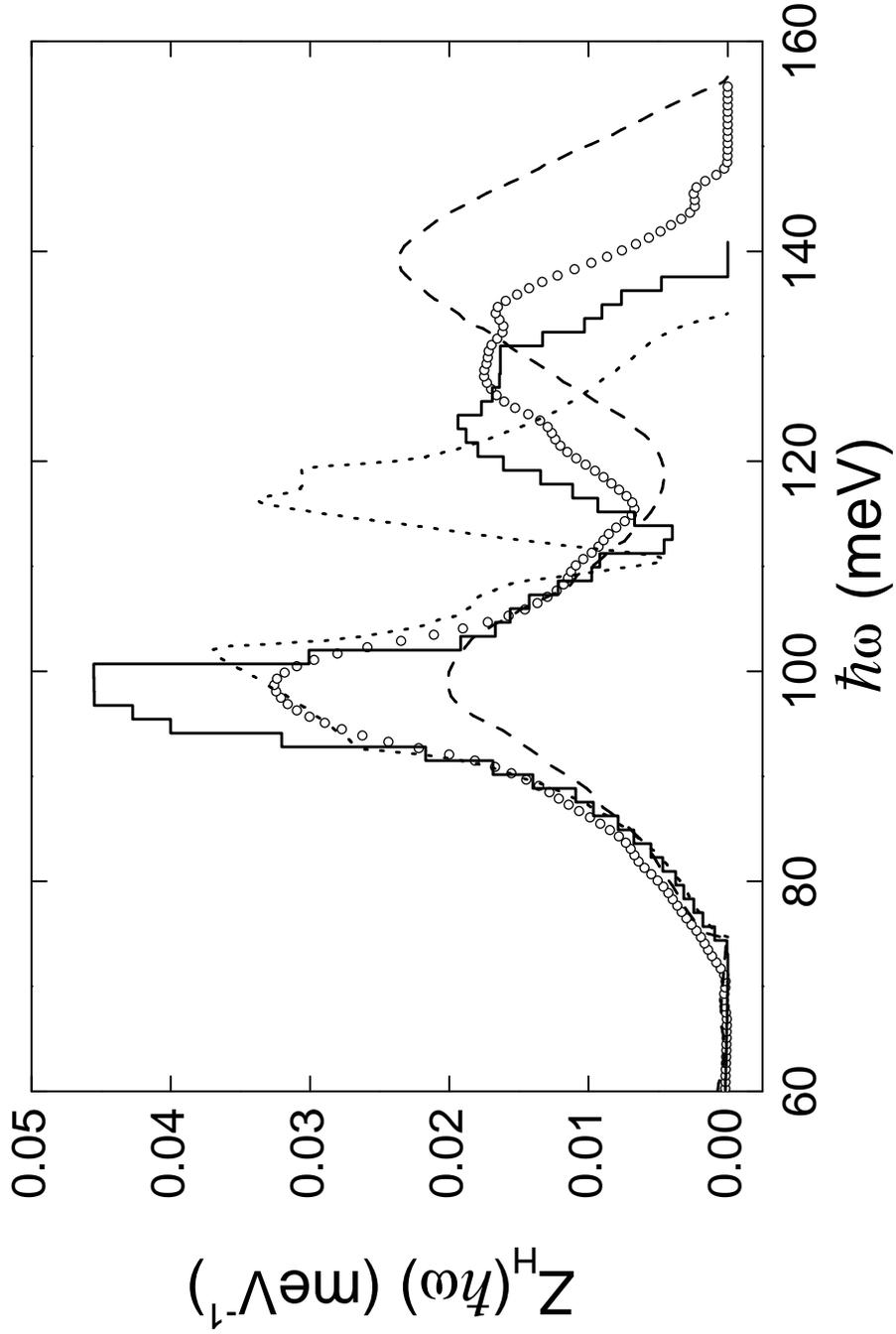}
\end{center}
\caption{
Hydrogen-projected density of phonon states in LiH. 
The experimentally-determined result is plotted as circles, the full
line represents a DDM13 lattice dynamics simulation,\protect\cite{Dyck} 
the dashed line a SM7 lattice dynamics simulation,\protect\cite{Verble} 
and the dotted line the old neutron measurement.\protect\cite{Zemlianov}
}  
\label{DOPS}
\end{figure}

The last stage before the extraction of the
H-DoPS was the  evaluation and the subtraction of the multiphonon
contribution, not totally negligible because of  the $Q$-values attained
by TOSCA in the $3$ meV $<\hbar \omega<$ $150$ meV range  (namely 2.8
\AA$^{-1}<Q<9.5$ \AA$^{-1}$). Processed LiH data, proportional to the
self inelastic structure factor \cite{Love} for the H ions,
$S_{s}(Q,\omega)$, were analyzed through an iterative self-consistent
procedure,\cite{muphip} aiming to extract  the one-phonon component of
$S_{s}(Q,\omega)$ (see Fig. \ref{MULT} (b)) and the hydrogen Debye-Waller factor, namely  
$S_{s,1}(Q,\omega)$ and $W(Q)$, respectively. Once again, all the
technicalities can be found in Ref. [\onlinecite{Colo}], while in the following only
the final equation of the procedure is reported.
From $S_{s,1}(Q,\omega)$ and $W(Q)$, given the isotropic nature of the LiH lattice,
$Z_{\rm H}(\omega)$ is simply worked out via:\cite{Love}
\begin{equation}
\label{densH}
Z_{\rm H}(\omega)=\exp \left( 2W(Q) \right)~S_{s,1}(Q,\omega) \frac{4 m_{\rm H} \omega}{\hbar Q^2} 
\left[ \coth \left( \frac{\hbar \omega}{2k_BT} \right) + 1 \right]^{-1}~,
\end{equation}
where $m_{\rm H}$ is the proton mass. The result for $Z_{\rm H}(\omega)$ is
plotted in Fig. \ref{DOPS}. 
Equation (\ref{densH}), and the other formulas \cite{Colo} used to work out $Z_{\rm H}(\omega)$, 
are formally exact only in the framework of the harmonic approximation, but they have a practical validity 
that is far more general, as explained by Glyde \cite{Glyde} in the context of solid helium.
From $Z_{\rm H}(\omega)$, making use of normal and Bose-corrected moment
sum rules,
\cite{Turchin} we are able to derive three important quantities related
to the hydrogen dynamics in this hydride, namely the H mean squared
displacement $\langle {\bf u}^2_{\rm H} \rangle $, the H mean kinetic energy
$\langle T_{\rm H} \rangle$, and the H Einstein frequency, $\Omega_{0,{\rm H}}$:
\begin{eqnarray}
\label{norbos}
\langle {\bf u}^2_{\rm H} \rangle &=&\frac{3 \hbar }{2 m_{\rm H}} \int_0^{\infty} d \omega \frac{Z_{\rm H}
(\omega)}{\omega} \coth \left( \frac{\hbar \omega}{2k_BT} \right)~, \nonumber \\
\langle T_{\rm H} \rangle&=& \frac{3 \hbar }{4} \int_0^{\infty} d \omega Z_{\rm H}(\omega)~\omega 
\coth \left( \frac{\hbar \omega}{2k_BT} \right)~, \nonumber \\
\Omega^2_{0,{\rm H}}&=& \int_0^{\infty} d \omega Z_{\rm H}(\omega)~ \omega^2~.
\end{eqnarray}
As for the first, we found: $\langle {\bf u}^2_{\rm H} \rangle=0.062(1)$ \AA$^2$, while for the
second: $\langle T_{\rm H} \rangle=80(1)$ meV, and finally for the last: 
$ \hbar \Omega_{0,{\rm H}}=109.2(9)$ meV. 
\section{Variational Monte Carlo Simulation}
We have studied the ground-state properties of solid LiH by means of the
variational Monte Carlo (VMC) method.\cite{Reynolds} VMC is a fully quantum 
approach which relies on the variational principle; it has been extensively 
used in the past in microscopic calculations of quantum crystals,\cite{Kalos} 
mainly $^4$He and $^3$He. Assuming only pair-wise interactions between the
different atoms in the crystal, the Hamiltonian describing LiH is:
\begin{equation}
H=-\frac{\hbar^2}{2 m_{\rm H}} \sum_{i=1}^{N_{\rm H}} \mbox{\bf
$\nabla$}_i^2 -\frac{\hbar^2}{2 m_{\rm Li}} \sum_{i=1}^{N_{\rm Li}} \mbox{\bf
$\nabla$}_i^2 + \sum_{i<j}^{N_{\rm H}} V^{({\rm H},{\rm H})}(r_{ij})
 + \sum_{i<j}^{N_{\rm Li}} V^{({\rm Li},{\rm Li})}(r_{ij})
 + \sum_{i,j}^{N_{\rm H},N_{\rm Li}} V^{({\rm H},{\rm Li})}(r_{ij})  \ ,
\label{hamiltonian}
\end{equation}
where $r_{ij}$ is the distance between the atoms composing an $i,j$ pair;
$m_{\rm Li}$ is the Li average atomic mass; $N_{\rm H}$ and $N_{\rm Li}$ stand for
the number of H and Li atoms, respectively; and $V^{({\rm H},{\rm H})}$,
$V^{({\rm Li},{\rm H})}$ and $V^{({\rm Li},{\rm Li})}$ represent the three 
pair-wise interaction potentials.
The variational principle states that for a given trial wave-function
$\Psi$, the expected value of $H$ is an upper bound to the ground-state
energy $E_0$:
\begin{equation}
\frac{\left \langle \Psi | H |\Psi \right \rangle}{\left \langle \Psi |
\Psi \right \rangle} = E \geq E_{0} \ .
\label{bound}
\end{equation} 
The multidimensional integral required in the calculation of Eq. (\ref{bound})
can not be evaluated exactly by analytical summation methods like, for example, the
hypernetted chain formalism.\cite{Rosati} However, a stochastic interpretation of this
integral is rather straightforward, and this is actually the task carried out by the
VMC method, with the only cost of some statistical noise. 
\par
A key point in the method is the search for a trial wave-function with
a sizeable overlap with the true ground-state wave-function $\Psi_0$. An
extensively tested model in quantum crystals is the Nosanow-Jastrow trial
wave-function.\cite{Glyde,Kalos} According to this description, the wave-function is written
as a product of a Jastrow factor $F$, accounting for the correlations induced
by the interatomic potentials, and a phase dependent term $\Phi$, which introduces
the crystal symmetries in the problem:
\begin{equation}
 \Psi = F \Phi  \ .
\label{nosanow}
\end{equation}
The Jastrow factor contains two-body correlation functions between the different
pairs $f^{({\rm H},{\rm H})}$, $f^{({\rm Li},{\rm Li})}$ and 
$f^{({\rm H},{\rm Li})}$:
\begin{equation}
F = \prod_{i<j}^{N_{\rm H}} f^{({\rm H},{\rm H})}(r_{ij}) 
\prod_{i<j}^{N_{\rm Li}} f^{({\rm Li},{\rm Li})}(r_{ij})
\prod_{i,j}^{N_{\rm H},N_{\rm Li}} f^{({\rm H},{\rm Li})}(r_{ij})  \ ,
\label{jastrow}
\end{equation}
and $\Phi$ localizes each particle around the lattice equilibrium sites of the crystal
phase ${\bf R}_i^\alpha$ through the functions $g^{({\rm H})}$ and $g^{({\rm Li})}$:
\begin{equation}
\Phi = \prod_{i}^{N_{\rm H}} g^{({\rm H})}(|{\bf r}_i - {\bf R}_i^{\rm H}|) 
\prod_{i}^{N_{\rm Li}} g^{({\rm Li})}(|{\bf r}_i - {\bf R}_i^{\rm Li}|) \ .
\label{phaseg}
\end{equation}
\par
The VMC simulation is carried out at a molar volume $v_0=10.059$ cm$^3$,
which is the experimentally estimated value at zero pressure and zero
temperature.\cite{Besson,Hama} 
The lattice constant, derived from the
experimental  molar volume $v_0$, is $a=4.0578$ \AA. The calculation is
worked out with a simulation cubic box containing 108 particles of each type
with periodic boundary conditions. We have checked that this number of
particles is large enough for practically eliminating size effects on the more
relevant quantities in the present study: the H mean kinetic energy, its mean squared
displacement around the lattice sites, and its Einstein frequency.
\par
As we have seen in the introductory section, LiH and the rest of hydrides and 
deuterides of light alkali metals are well described as ionic crystals. This nearly 
perfect ionic bonding allows for a model in which ions H$^-$ and Li$^+$, 
both with a 1s$^2$ electronic configuration, are occupying the lattice sites of 
the crystal. The present simulation
relies on this model, assuming rigid closed-shell ions interacting via
central interatomic potentials. The overlap repulsion potential between the
ions is taken from the semi-empirical Born-Mayer interaction,\cite{Born}
and van der Waals attractive terms are also included to deal with 
polarizability effects.
According to this general scheme, as previously mentioned, Haque and Islam
\cite{Haque} proposed their pair potential $V_{\rm HI}^{(\alpha,\beta)}(r)$:
\begin{equation}
V_{\rm HI}^{(\alpha,\beta)}(r) = B_{\alpha \beta} \exp{ \left( - A_ {\alpha
\beta} r \right) } - \frac{C_{\alpha \beta}}{r^6}  \ ,
\label{haque}  
\end{equation}      
with $\{\alpha$, $\beta \}$ = Li, H. The set of parameters entering $V_{\rm
HI}^{(\alpha,\beta)}(r)$ is reported in Tab. I. In order to evaluate the 
influence of the interatomic potentials in our results we have also used the 
model proposed by Sangster and Atwood,\cite{Sangster} 
$V_{\rm SA}^{(\alpha,\beta)}(r)$:
\begin{equation}
V_{\rm SA}^{(\alpha,\beta)}(r) = B_{\alpha \beta} \exp{ \left[  A_ {\alpha
\beta}  \left( E_{\alpha \beta} + F_{\alpha \beta} - r \right) \right] } - 
\frac{C_{\alpha \beta}}{r^6} - \frac{D_{\alpha \beta}}{r^8}  \ ,
\label{sangster}  
\end{equation}   
which incorporates a dipole-quadrupole attractive term. The parameters for
LiH, which are taken from Hussain and Sangster \cite{Huss} (see also Sect. I), 
are reported in Tab. II.
\par
The main task in a variational approach like the present one is to seek for
a good trial wave-function $\Psi$ (Eqs. (\ref{nosanow}-\ref{phaseg})). 
Like in solid helium calculations,\cite{Kalos} we have chosen analytical 
two-body $f^{(\alpha,\beta)}(r)$ (see Eq. (\ref{jastrow})) and one-body 
$g^{(\alpha)}(r)$ (see Eq. (\ref{phaseg})) correlation factors with a set 
of free parameters to be optimized.
In particular, $f^{(\alpha,\beta)}(r)$ is of McMillan type:\cite{Mcmillan}
\begin{equation}
f^{(\alpha,\beta)}(r) = \exp{ \left[ -\frac{1}{2} \left( \frac{b_{\alpha
\beta}}{r} \right)^5 \right] } \ ,
\label{mcmillan}
\end{equation}
and the specific phase factor is a Gaussian centered on the sites of the
crystal lattice:
\begin{equation}
g^{(\alpha)}(r)= \exp{  \left( - \frac{1}{2} \, c_{\alpha} r^2 \right) } \ .
\label{onebody}
\end{equation}
In the optimization search the absolute minimum of the internal energy
$\langle H \rangle_{\Psi}$ is looked for. However, the number of parameters
is large enough to make this optimization rather difficult. In order to
discern between local minima of similar quality we have also calculated the
energy of the solid including Coulomb contributions, considering the ions as
point-like particles. Under this criterion, the final parameter set
is the one which simultaneously minimizes the short-range energy and the
total energy including Coulomb contributions. The values obtained, which
are the same for the two short-range potentials $V_{\rm HI}$ and $V_{\rm
SA}$, are: $c_{\rm Li}=150$ \AA$^{-2}$, $c_{\rm H}=15$ \AA$^{-2}$, 
$b_{{\rm Li} {\rm H}}=2.0$ \AA, $b_{{\rm Li} {\rm Li}}=1.5$ \AA, and 
$b_{{\rm H} {\rm H}}=1.0$ \AA.  Among the three $b_{\alpha \beta}$ Jastrow
parameters, the most important is the cross one ($b_{{\rm Li} {\rm H}}$),
since the first neighbors of a H$^-$ ion are Li$^+$, and vice versa. It is
worth noticing that the optimal values for $c_{\alpha}$ reflect the
difference in the degree of localization around the sites 
between the two ions due to their
significantly different masses, $c_{\rm H} << c_{\rm Li}$. Including the Coulomb potential
in the energy, the energies per particle are -5.07(2) eV and
-5.32(2) eV for the $V_{\rm HI}$ and  $V_{\rm SA}$ short-range potentials,
respectively.
\par
Information on the spatial structure of the solid can be drawn from the
two-body radial distribution functions $g^{(\alpha,\beta)}(r)$. In  Fig. 3,
results for the three components  $g^{(\alpha,\beta)}(r)$ are shown. As
expected, the location of the peaks follows the inter-particle pattern
imposed by the lattice: each ion is surrounded by ions
of opposite sign and ions of equal sign are distributed with the same
periodicity. The major mobility of H$^-$ with respect to Li$^+$ is also observed by
comparing the height and the spreading around the sites of $g^{({\rm Li},{\rm
Li})}(r)$, on one hand, and  $g^{({\rm H},{\rm H})}(r)$ and to a lesser
extend $g^{({\rm Li},{\rm H})}(r)$, on the other. 
\par
A structural quantity which can be directly compared with the present
experimental data is the mean squared displacement of the H$^-$ ions, 
$\langle {\bf u}^2_{\rm H} \rangle$:
\begin{equation}
\langle {\bf u}^2_{\rm H} \rangle = \frac{1}{N_{\rm H}} \, \left \langle
\sum_{i=1}^{N_{\rm H}} \left( {\bf r}_i - {\bf R}_i^{\rm H} \right )^2
\right \rangle  \ .
\label{meandis}
\end{equation} 
Using the trial wave-function quoted above, the VMC result is $\langle
{\bf u}^2_{\rm H} \rangle =0.074(2) $\AA$^2$. The resulting Lindemann ratio:
\begin{equation}
\gamma_{\rm H}=\frac{2 \sqrt{ \langle {\bf u}^2_{\rm H} \rangle }}{a} \ ,
\label{lindemann} 
\end{equation}  
is 0.134. Additional insight
on the spatial localization of H$^-$ can be obtained by calculating the
H$^-$ density profile $u_{\rm H} (r)$, with $r$ being the distance between the ion and
its site. The function $u_{\rm H} (r)$ is shown in Fig. 4; it is very well
parameterized by a Gaussian with the VMC expected value $\langle {\bf u}^2_{\rm H} \rangle 
=0.074(1)$ \AA$^2$ (see the solid line in the same figure). 
\par
The kinetic energy per particle of H$^-$ is one of the partial
contributions to the total energy of
the solid,
which is evaluated at each step of the VMC simulation. It is the expected value of
the operator:
\begin{equation}
\langle T_{\rm H} \rangle = -\frac{\hbar^2}{2 m_{\rm H} N_{\rm H}} \left \langle \sum_{i=1}^{N_{\rm H}} 
\frac{\mbox{\bf $\nabla$}_i^2 \Psi}{\Psi } \right \rangle \ ,
\label{ekinmc} 
\end{equation}
with configuration points generated according to the probability
distribution function $|\Psi|^2$. As in the estimation of $\langle
{\bf u}^2_{\rm H} \rangle$,
the H$^-$ kinetic energy is the same for the two interatomic potentials
(Eqs. \ref{haque},\ref{sangster}) since the optimization procedure has led
to the same variational wave-function. The result obtained is $\langle
T_{\rm H} \rangle= 84(1)$meV.
\par
The third physical quantity evaluated in the present experiment, the Einstein
frequency $\Omega_{0,{\rm H}}$, can be  calculated in a
VMC simulation through its proper definition:
\begin{equation}
\Omega_{0,{\rm H}}^{2} =
\frac{1}{3 m_{\rm H} N_{\rm H}} \left \langle \sum_{i=1}^{N_{\rm H}}  
\mbox{\bf $\nabla$}_i^2
V_{\rm H}({\bf r}_{i}) \right \rangle  \ ,
\label{omega2}
\end{equation}
where the expected value of $ \mbox{\bf $\nabla$}_i^2 V_{\rm H}(r)$ is calculated over
the configurations generated
by the probability distribution function $|\Psi|^2$, and $V_{\rm
H}(r)$ is the potential felt by an H$^-$ ion:
\begin{equation}
V_{\rm H}({\bf r}_{i}) = \frac{1}{2} \sum_{j \neq i}^{N_{\rm H}} V^{({\rm H},{\rm H})}
(r_{ij})+ \frac{1}{2} \sum_{j=1}^{N_{\rm Li}} V^{({\rm H},{\rm Li})}(r_{ij}) \ .
\label{pothidro}
\end{equation}
The result obtained applying Eq. (\ref{omega2}) is $\hbar \Omega_{0,{\rm
H}} = 110.3(5)$ meV.
\section{Discussion}
The aim of this discussion section is threefold:  1) critically analyzing
various dynamical quantities derived  from the scientific literature with
respect to the present results obtained through neutron scattering 
(Subsect. A);  2) comparing the outputs of the experimental and theoretical
approaches
presently used, namely IINS and VMC simulations  (Subsect. B);  3)
finally, shedding some light on the important point of the possible
quantum anharmonic effects on the H$^-$ dynamics in low-temperature LiH
(Subsect. C).
\subsection{Analysis of the H-DoPS evaluations}
A comparison among the various determinations (both experimental and
simulated) of the hydrogen-projected density  of phonon states could be
finally established at this stage. In Fig. \ref{DOPS}, four H-DoPS
estimates have been plotted  together in the frequency region concerning
the two optical bands ($65$ meV $< \hbar \omega<160$ meV), namely the
translational optical (TO) and the longitudinal optical (LO), which
actually contain more than $97$\% of the  total $Z_{\rm H}(\omega)$ area.
\cite{Dyck} The present IINS experimental result is plotted as circles,
the full line represents  the DDM13 lattice dynamics simulation,
\cite{Dyck} the dashed line is the SM7 lattice dynamics simulation,
\cite{Verble}  and the dotted line stands for the old IINS measurement.
\cite{Zemlianov} As a preliminary comment, one can easily observe the
existence of a fair general agreement among all the four  $Z_{\rm
H}(\omega)$ in the TO range, at least as far as the peak position is
concerned. On the contrary, the  LO region looks much more uncertain, the
peak centroid varying from 115 meV up to 140 meV. The reason for such a 
behavior is easily understandable for SM7 and DDM13 data: these lattice
dynamics calculations made use of the some parameters (7 and 13,
respectively) derived from a fit of the same LiD dispersion curves
measured by Verble \cite{Verble} (for a detailed comparative discussion on
the differences between the SM7 and DDM13 H-DoPS calculations see Ref.
[\onlinecite{Dyck}]). By a simple inspection of these experimental
dispersion curves, it is clear  that the LO neutron groups are really few
(four values plus one infra-red measurement at the $\Gamma$ point). 
However the disagreement between the present IINS $Z_{\rm H}(\omega)$ and
the old one is difficult to explain, so that  we are inclined to think
that these discrepancies are due to experimental imperfections in the data
analysis of the latter  (e.g. multiple scattering or multiphonon scattering
subtraction). Selecting the two most recent experimental and  simulated
$Z_{\rm H}(\omega)$, namely the present IINS and the DDM13 estimates, we
can observe an overall  semi-quantitative agreement, the main
discrepancies being concentrated in two regions: at low energy, in the
onset of  TO band ($65$ meV $< \hbar \omega < 90$ meV), and in the LO band
as a whole ($112$ meV $< \hbar \omega < 145$ meV). As for the latter, a
simple energy shift of $4.5$ meV seems sufficient to largely reconcile
IINS and DDM13, while in the  former case, neutron data appear somehow
broader than lattice dynamical ones (IINS FWHM being about $5.1$ meV
larger  than DDM13 FWHM).  Considering the TOSCA-II energy resolution in
this region ($\Delta \hbar \omega \simeq 1.6$ meV), a simple explanation 
based only on experimental effects can be easily discarded. However, as
pointed out by Izyumov and Chernoplekov for other  hydrides,
\cite{Izyumov} such a broadening of the H-DoPS TO bands might be the mark
of the hydrogen anharmonic dynamics  in LiH, through a finite phonon
life-time. In this respect, more will be said in Subsect. C.
\subsection{Comparison between IINS and VMC results}
Since VMC, as we have seen in Sect. III, is a ground-state method, $Z_{\rm
H}(\omega)$ can not be directly evaluated. However, through the aforementioned
normal and  Bose-corrected moment sum rules in Eqs. (\ref{norbos}), it was
possible to describe the main features of the H-DoPS via $\langle {\bf
u}^2_{\rm H} \rangle $, $\langle T_{\rm H} \rangle$ and $\Omega_{0,{\rm
H}}$, which are equilibrium quantities calculated by the VMC code
(see also Sect. III).  Before proceeding with this comparison, it is worth
noticing that the VMC calculation is performed at zero temperature, 
whereas the measure is accomplished at $T=20.1(1)$ K. However, thermal effects
are negligible since the Debye  temperature of LiH is approximately $1100$
K,\cite{Islam} and therefore the measured system can be certainly 
considered in its ground state, at least as far as the H$^-$ ion dynamics
is concerned. This assumption can be easily proved by calculating (always
from the experimental $Z_{\rm H}(\omega)$) the zero-point values of the H
mean squared displacement and mean kinetic energy, setting $T=0$: $\langle
{\bf u}^2_{\rm H} \rangle(T=0)=0.062(1)$\AA$^2$ and  $\langle T_{\rm H}
\rangle(T=0)=$80(1) meV, identical within the errors to the values
estimated at $T=20.1$ K 
in Sect. II.
\par
Going back to VMC, one can notice a value of the zero-point H mean squared
displacement ($0.074(2)$ \AA$^2$,  as in Sect. III) slightly higher than
the IINS experimental measure. In addition these two figure have to be
compared  to the most recent neutron diffraction estimate by Vidal and
Vidal-Valat:\cite{Vid}  $\langle {\bf u}^2_{\rm H} \rangle=$0.0557(6)
\AA$^2$ (extrapolated at $T=20$ K by the present authors from the
original  data in the temperature range $93$ K$-293$ K), which appears
close but still discrepant from the VMC and IINS findings.  However,  it
has to be pointed out that a previous room-temperature diffraction result
by Calder {\it et al.},\cite{Calder}  $\langle {\bf u}^2_{\rm H}
\rangle=$0.068(1) \AA$^2$, seems to exhibit a similar trend if compared to
the Vidal and Vidal-Valat's figure in the same conditions: $\langle {\bf
u}^2_{\rm H} \rangle=$0.0650(6) \AA$^2$.  On the other hand, for the other
two aforementioned physical quantities, the agreement between IINS 
($\langle T_{\rm H} \rangle=80(1)$ meV, $\hbar \Omega_{0,{\rm
H}}=109.2(9)$ meV) and VMC ($\langle T_{\rm H} \rangle=84(1)$ meV, $\hbar
\Omega_{0,{\rm H}}=110.3(5)$ meV) is much more satisfactory, confirming
the validity of our combined IINS-VMC method.
\par
An interesting test on the obtained results can be accomplished in the
framework of an approximate estimation known  as the Self Consistent
Average Phonon (SCAP) formalism.\cite{Shukla,Paskin} The SCAP approach
relies on the well-known Self Consistent Phonon (SCP) method, but
replacing the sums of functions of the phonon frequencies by  appropriate
functions of an average-phonon frequency. Results \cite{Paskin} for
quasi-harmonic and harmonic solids like  Ne, Kr, and Xe obtained using
SCAP have shown good agreement with experimental data. The application of
this formalism to  quantum crystals seems however more uncertain due to
the relevant increase of anharmonicity. As LiH seems to be a  quantum
crystal, but with less anharmonicity than for example $^4$He, SCAP can
somehow help in the present study.  Normally SCAP is used to evaluate
various physical quantities in an iterative way, employing lattice
parameters and  interatomic potentials only.\cite{Shukla} Here, on the
contrary, the method will be applied in one single step, starting from
``exact" values of $\Omega_{0,{\rm H}}$. To this end, we have calculated
via SCAP (at $T=0$) the  H mean squared displacement and mean kinetic
energy using the relations: 
\begin{eqnarray} 
\langle {\bf u}_{\rm H}^{2}
\rangle^{\rm (SCAP)} & =& \frac{3 \hbar}{2 m_{\rm H} \Omega_{0,{\rm H}}}
~, \nonumber \\ \langle T_{\rm H} \rangle^{\rm (SCAP)} & = & \frac{1}{2} ~
m_{\rm H} \langle {\bf u}_{\rm H}^{2} \rangle^{\rm (SCAP)}  \Omega_{0,{\rm
H}}^2 = \frac{3}{4} \hbar \Omega_{0,{\rm H}}~. 
\end{eqnarray} 
The results
obtained through this approximation are:  $\langle {\bf u}^2_{\rm H}
\rangle^{\rm (SCAP-VMC)}=0.0564(3)$ \AA$^2$,  $\langle {\bf u}^2_{\rm H}
\rangle^{\rm (SCAP-IINS)}=0.0570(5)$ \AA$^2$,  $\langle T_{\rm H}
\rangle^{(\rm SCAP-VMC)}=82.7(4)$ meV, and $\langle T_{\rm H} \rangle^{\rm
(SCAP-IINS)}=81.9(7)$ meV. By comparing these approximated values with the
microscopic ones, one realizes that the H mean kinetic energy
values come out very close, but the SCAP values of the mean squared 
displacement are significantly smaller, actually not far from the neutron
diffraction result by Vidal and Vidal-Valat.\cite{Vid} The physical
meaning of these results is straightforward: both SCAP equations are exact
at $T=0$ in presence of a purely harmonic Einstein solid (i.e. if $Z_{\rm
H}(\omega)=$$\delta \left(\omega-\Omega_{0,{\rm H}} \right)$ ). But  if
the solid system exhibits a broader H-DoPS, then $\langle {\bf u}_{\rm
H}^{2} \rangle^{\rm (SCAP)}$ comes out rather underestimated, since
$\Omega_{0,{\rm H}}^2$ is exactly computed stressing the high-frequency
part  of $Z_{\rm H}(\omega)$ via the integrand factor $\omega^2$ (as in
Eqs. (\ref{norbos})). In this way, one somehow corrects for this bias by
expressing the zero-point mean kinetic energy as the product of $\langle
{\bf u}^2_{\rm H} \rangle^{\rm (SCAP)} $ times $\Omega_{0,{\rm H}}^2$,
because $\langle T_{\rm H} \rangle$ would be exactly evaluated via the
integrand factor $\omega$  (see Eqs. (\ref{norbos})), which still stress
the high-frequency part of  $Z_{\rm H}(\omega)$, but less than
$\Omega_{0,{\rm H}}^2$.  Nevertheless, it is quite remarkable the good accuracy achieved
by this relatively simple approach in evaluating the mean kinetic energy,
probably beyond what is a priori expected for a system with a possible
quantum-crystal character like LiH.
\subsection{Possible quantum anharmonic effecs in LiH}
Given the relatively large value of the H Lindemann ratio in LiH 
($\gamma_{H}\simeq(0.12-0.13)$ as seen above), it is natural to inquire on the
low-temperature anharmonic effects in the H$^-$ ion dynamics. In this
respect our tools are well suited, since VMC is a microscopic quantum
simulation, assuming neither the harmonic approximation like the usual
lattice dynamic calculations, nor the semi-classical treatment of the
particle motion like the molecular dynamics approach.\cite{semicl} The
method applied to test the existence of possible quantum anharmonic effects
in H dynamics (at $T=0$) was simply devised running the same VMC  code in a
``harmonic way", i.e., replacing for every atomic pair $i,j$ the  exact pair
potential value $V^{(\alpha,\beta)}(r_{ij})$ by:\cite{Wallace} 
\begin{equation}
V^{(\alpha,\beta)}(r_{ij}) \simeq V^{(\alpha,\beta)}(r_{0,ij}) + 
\frac{1}{2} \left( {\bf u}_i -
{\bf u}_j \right)^{\rm T} \left( \frac{ \partial ^2 V^{(\alpha,\beta)}
(r_{ij}) }{ \partial
{\bf r}_{ij} \partial {\bf r}_{ij} } \right)_{ r_{ij}= r_{0,ij}}  \left(
{\bf u}_i - {\bf u}_j \right)~, 
\label{anharmonic}
\end{equation}   
where ${\bf r}_i$ and
${\bf u}_i$ are the instantaneous position of an $i$ atom and its
displacement from the equilibrium position, respectively, while ${\bf
r}_{ij}$ stands for the vector ${\bf r}_i-{\bf r}_j$, and $r_{0,ij}$ is
the equilibrium separation of  an atomic pair $i,j$. It is worth noting
that both the static potential energy   $V^{(\alpha,\beta)}(r_{0,ij})$ and the
Hessian components are all calculated only once per each  atomic pair 
$i,j$, since they depend only on the equilibrium distance $r_{0,ij}$.
However  the ``harmonic" results from VMC did not show any significant
difference (within their uncertainties) from the ``exact" ones (see above
in Subsect. B), proving that, at least in the  framework of the
semi-empirical pair potentials employed, quantum anharmonic effects are
totally  negligible in the evaluation of $\langle {\bf u}^2_{\rm H}
\rangle $ and  $\langle T_{\rm H} \rangle$, even at $T=0$.   
\section{Conclusions and perspectives}
Dynamical properties of solid LiH at low temperature have been studied
using incoherent inelastic neutron scattering with higher accuracy than 
in previous measurements. 
The analysis of the scattering data has allowed for the extraction of a 
reliable hydrogen-projected density of phonon states.
From this physical quantity it was possible to estimate
three relevant quantities intimately related to the microscopic
dynamics of H in the solid: its mean squared displacement, mean kinetic
energy, and  Einstein frequency. Apart from the intrinsic interest in an
accurate quantitative determination of these magnitudes, we have
tried to shed some light on two fundamental questions on the physical nature of
LiH, i.e. its quantum character and the degree of anharmonicity in the H
dynamics. To this
end, and also to make a direct comparison with theory, we have carried out
a quantum microscopic calculation of the same three quantities quoted above
using VMC.   
\par
The consideration of solid LiH as a quantum solid seems already justified by
its Lindemann ratio which is smaller than the two paradigms, $^4$He and $^3$He,
but still appreciably larger than the common values in classical solids.
From a theoretical viewpoint, it supposes the unavoidable introduction of
at least two-body correlations to account correctly for its ground-state
properties. We have verified using VMC that this feature also holds in
solid LiH. 
\par
The degree of anharmonicity in H dynamics has been established by comparing the full VMC
calculation with another one in which the real interatomic potentials have
been substituted by harmonic approximations (Eq. (\ref{anharmonic})). As
commented in the previous Section, both VMC simulations generate identical
results and then possible anharmonic effects in H are not observed. This
situation is different from the one observed in solid $^4$He where quantum
character and anharmonicity appear together. Apart from the appreciable
difference between both systems looking at their respective Lindemann ratios,
a relevant feature that can help to understand the absence of anharmonicity
of H 
in LiH is the significant difference between the interatomic potentials at
short distances in both systems. Helium atoms interact with a
hard core of Lennard-Jones type whereas the short-range interaction between
the components of the mixture LiH is much softer (exponential type) according to the
Born-Mayer model. 
\par
The agreement achieved in the present work between the neutron scattering
data and the VMC calculation is remarkably good and probably better than
what could be initially expected from the use of semi-empirical
interactions. The VMC predictions for the H kinetic energy and Einstein
frequency coincide within error bars with the experimental measures. On the
contrary, the H mean square displacement is about $15$\% larger. This points
to probable inaccuracies of the model potentials, in particular, to core sizes 
smaller than the real ones. 
In this respect, it has also to be said that these semi-empirical potentials
have been proposed and evaluated in order to reproduce various LiH properties 
(e.g. lattice constant, bulk modulus, reststrahl frequency, etc.)\cite{Huss,Haque,Sangster} 
in the framework of the standard (i.e. harmonic) lattice dynamics.
Thus, the {\it ab initio} calculation of more realistic pair interactions 
could help enormously to improve a microscopic description of LiH and justify for 
the future the use of techniques beyond VMC, like the diffusion Monte Carlo or 
path integral Monte Carlo methods.
\section*{Acknowledgements}
The experimental part of this work has been financially supported by C.N.R. (Italy).
Dr. A. J. Ramirez-Cuesta (Rutherford Appleton Laboratory, U.K.) is gratefully 
thanked by some of the authors (D. C. and M. Z.) for his skillful support.
J. B. and C. C. acknowledge partial financial support from DGI (Spain) 
Grant No. BFM2002-00466 and
Generalitat de Catalunya Grant No. 2001SGR-00222.
%

%
\pagebreak

\begin{table}
\centering
\begin{ruledtabular}
\begin{tabular}{ccccc}
$\alpha$  &  $\beta$  &   $A_{\alpha \beta}$(\AA$^{-1}$)  &  $B_{\alpha
\beta}$(eV)   &  $ C_{\alpha \beta}$(eV\AA$^{6}$)   \\
\tableline
Li  &  Li     &	  7.3314     &   1153.80    &   0.0        \\                            
Li  &  H      &	  3.1000     &   187.29     &   0.0        \\                        
H   &  H      &   5.5411       &   915.50     &   4.986      
\end{tabular}
\end{ruledtabular}
\caption{
Parameters of the $V_{\rm HI}^{(\alpha,\beta)}(r)$ potential by Haque and
Islam.\protect\cite{Haque}
}
\end{table}

\begin{table}
\centering
\begin{ruledtabular}
\begin{tabular}{cccccccc}
$\alpha$  &  $\beta$  & $A _{\alpha \beta}$(\AA$^{-1}$) &  $B_{\alpha \beta}$(eV) 
& $E_{\alpha \beta}$(\AA) & $F_{\alpha \beta}$(\AA) & $C_{\alpha
\beta}$(eV\AA$^{6}$) & $D_{\alpha \beta}$(eV\AA$^{8}$) \\
\tableline
Li  & Li &  38.48731  & 17.8000 & 0.2226 & 0.2226 & 0.0549073 & 0.0216285 \\ 
Li  & H  &  3.784068  & 0.17675 & 0.2226 & 1.9000 & 0.9568660 & 2.0013052 \\ 
H   & H  &  2.608380  & 0.01422 & 1.9000 & 1.9000 & 48.872096 & 185.18284 
\end{tabular}
\end{ruledtabular}
\caption{
Parameters of the $V_{\rm SA}^{(\alpha,\beta)}(r)$ potential proposed by Sangster 
and Atwood,\protect\cite{Sangster} estimated by Hussain and
Sangster.\protect\cite{Huss}
}
\end{table}
\begin{figure}
\begin{center}
\epsfxsize=35pc
\epsfbox{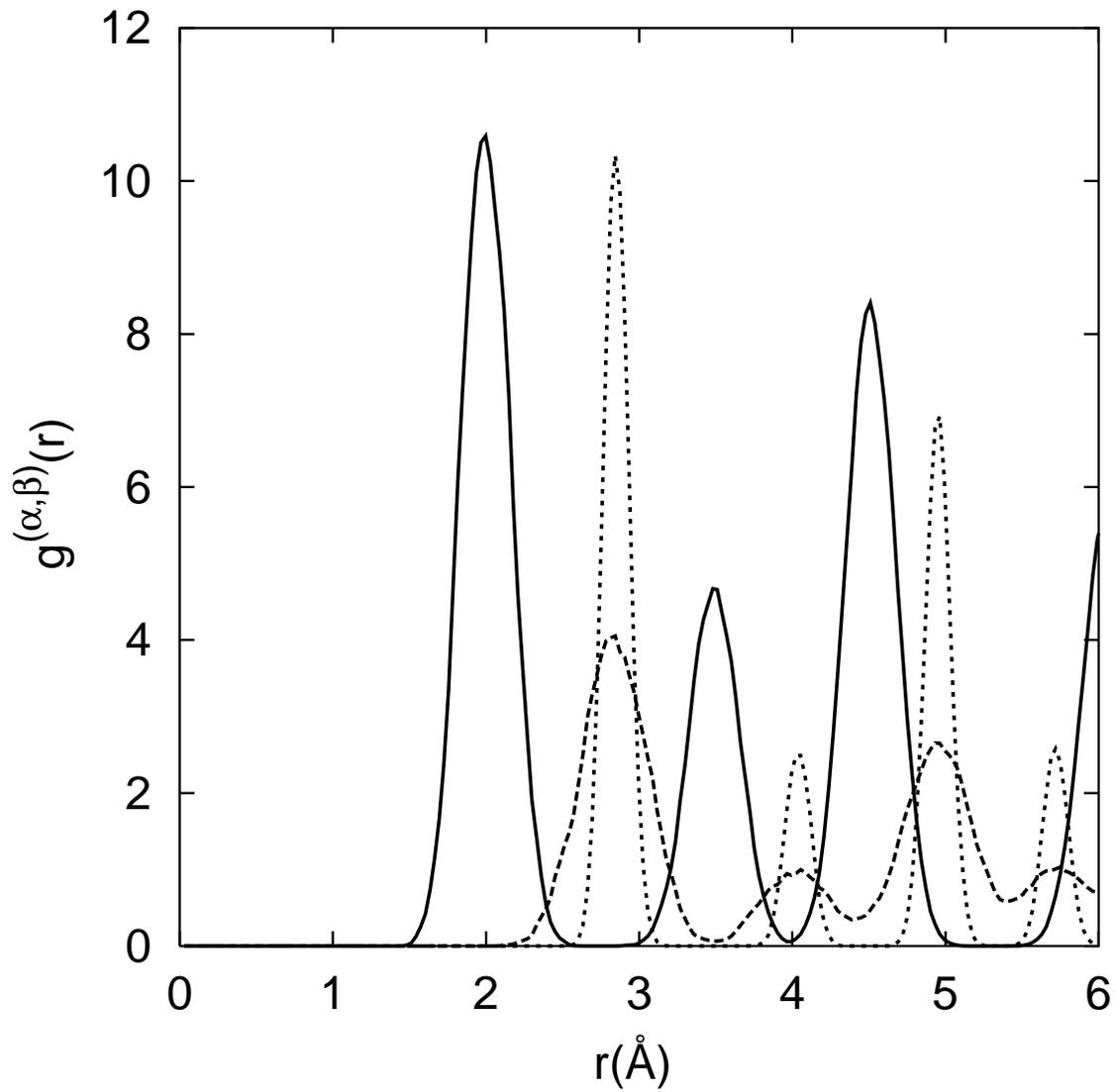}
\end{center}
\caption{
Two-body radial distribution functions in LiH: $g^{({\rm H},{\rm Li})}(r)$,
solid line; $g^{({\rm H},{\rm H})}(r)$, dashed line; $g^{({\rm Li},{\rm
Li})}(r)$, dotted line.
}
\end{figure}
\begin{figure}
\begin{center}
\epsfxsize=35pc
\epsfbox{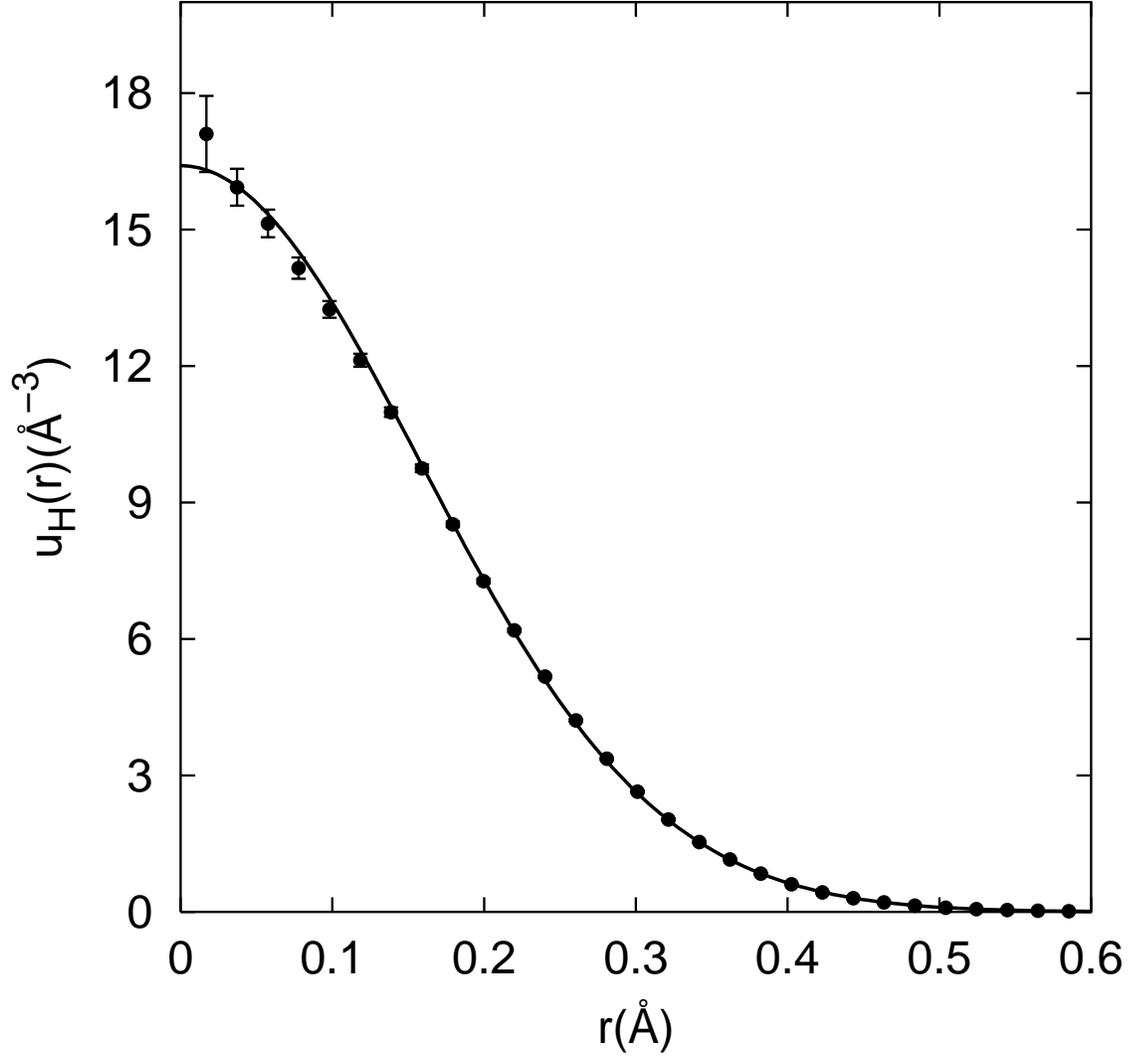}
\end{center}
\caption{
Variational Monte Carlo result for the H$^-$ density profile, $u_{\rm H}(r)$, in LiH. The solid line
corresponds to a Gaussian with a mean squared displacement equal to the variational Monte Carlo
value $\langle {\bf u}^2_{\rm H} \rangle=0.074$ \AA$^2$. 
}
\end{figure}
\end{document}